Australian Information Warfare and Security Conference

Conferences, Symposia and Campus Events

2015

# A hybrid feature selection for network intrusion detection systems: Central points

Nour Moustafa
*University of New South Wales*, nour.abdelhameed@student.adfa.edu.au

Jill Slay
*University of New South Wales*, j.slay@adfa.edu.au

# A HYBRID FEATURE SELECTION FOR NETWORK INTRUSION DETECTION SYSTEMS: CENTRAL POINTS AND ASSOCIATION RULES

Nour Moustafa, Jill Slay
University of New South Wales, Canberra, Australia
Nour.AbdElhameed@student.adfa.edu.au, j.slay@adfa.edu.au

**Abstract**
*Network intrusion detection systems are an active area of research to identify threats that face computer networks. Network packets comprise of high dimensions which require huge effort to be examined effectively. As these dimensions contain some irrelevant features, they cause a high False Alarm Rate (FAR). In this paper, we propose a hybrid method as a feature selection, based on the central points of attribute values and an Association Rule Mining algorithm to decrease the FAR. This algorithm is designed to be implemented in a short processing time, due to its dependency on the central points of feature values with partitioning data records into equal parts. This algorithm is applied on the UNSW-NB15 and the NSLKDD data sets to adopt the highest ranked features. Some existing techniques are used to measure the accuracy and FAR. The experimental results show the proposed model is able to improve the accuracy and decrease the FAR. Furthermore, its processing time is extremely short.*

## INTRODUCTION

Attackers attempt to breach computer networks to steal valuable information or disrupt computer resources. A Network Intrusion Detection System (NIDS) is a powerful defence mechanism that can defend against hostile threats of attackers (Lee, Stolfo, & Mok, 1999). NIDS methodologies can be classified as *Misuse Detection (MD)* and *Anomaly Detection (AD)* (Moustafa & Slay, 2015a; Valdes & Anderson, 1995; Vigna & Kemmerer, 1999). MD uses signatures of existing attacks to define known attacks. AD creates a normal profile of activities, and any strong deviations from this profile are considered as an attack. MD reduces False Alarm Rates (FAR), though it detects existing attacks only. Conversely, AD increases FAR, though it detects novel attacks. As a result, AD has become a critical point of research to reduce the FAR and increase the detection rate with identifying both existing and new attacks (Aziz, Azar, Hassanien, & Hanafy, 2014; Garcia-Teodoro, Diaz-Verdejo, Maciá-Fernández, & Vázquez, 2009).

Network packets consist of multiple features, due to the diversity of involved protocols and services. Some of these features are redundant or irrelevant. It can be observed that the redundant features are a major reason of increasing the FAR and decreasing the detection rate. A Feature Selection (FS) is a method of adopting the relevant features in a data set. The FS also reduces the computational time to implement an online NIDS. The reliable NIDS depends on removing noisy and redundant features (Hall, 1999).

In this study, we suggest a hybrid method of the Central Points of attribute values and an Association Rule Mining (ARM) technique. First, the Central Points of attribute values (CP) method mean computing the most repeated values of each attribute either attribute type is numerical or categorical. Second, the ARM model was developed to generate the highest correlated values of observations in a data set (Agrawal, Imieliński, & Swami, 1993; Zhang & Zhang, 2002), but we customise it as a feature selection. To establish an effective and reliable AD, the CP method computes the highest redundant values based on partitioning data observations into equal parts.as a consequence, this method reduces the processing time of executing the ARM technique.

## RELATED WORK AND BACKGROUND

The goal of ARM makes the strongest itemsets of features via computing support and confidence of rules (Agrawal et al., 1993; Lee & Stolfo, 1998; Yanyan & Yuan, 2010; Zhang & Zhang, 2002). Many studies have been suggested to apply ARM techniques in NIDSs. (Agrawal et al., 1993) proposed that program implementations and user activities can be correlated by using an ARM to create the most frequent attributes. (Lee & Stolfo, 1998) utilised the ARM to elicit rules for system audit data in order to build a normal user profile, any deviation from the profile established new rules. However, in both these studies, the computational time of the ARM is extremely high. (Yanyan & Yuan, 2010) designed a partition-based ARM model. The model was customised to scan the training set twice. During the first scan, the data set was divided into several partitions to be executed easily in memory, whereas during the second scan, the itemsets of the training set were established. Though, the complexity of this algorithm is highly expensive.

(Su, Yu, & Lin, 2009) developed an incremental fuzzy ARM technique. They used a linked-list method to store all candidate itemsets and their support in memory. The main disadvantage of this algorithm requires a large-size memory to store all candidate itemsets. (Nath, Bhattacharyya, & Ghosh, 2011) discussed a survey of existing feature selection approaches based on ARM methods. Some of the approaches used single objective functions, while others used with multi-objective. The results showed that the multi-objective ARM can be used to solve several real datasets. This study is partially related to our work to apply the ARM as feature selection.

## DESCRIPTION OF THE NSLKDD DATA SET

A NSLKDD data set is an enhanced version of the KDD99 data set ("NSLKDD," 2015). This data set has four attack categories: DoS, U2R, R2L and probe; and contains 42 features. In the NSLKDD data set, three major problems were addressed. First, the repeated observations in training and testing sets were detached to exclude biasing classification techniques towards the most frequent observations. Second, the training set and testing set were generated by selecting observations from different parts of the original KDD99 data set. Finally, the imbalanced of observations in each class either in the training set or testing set were solved to decrease the FAR. Table 1 shows the distribution of attack and normal records in the NSLKDD data set for the training and testing sets.

*Table 1: NSLKDD Data Set Distribution*

| Category | Training set | Testing set |
|---|---|---|
| DoS | 45,927 | 7,458 |
| U2R | 52 | 67 |
| R2L | 995 | 2,887 |
| Probe | 11,656 | 2,422 |
| Normal | 67,343 | 9,710 |
| **Total Records** | 125,973 | 22,544 |

The NSLKDD data set has disadvantages which can negatively affect the fidelity of NIDS evaluation. First, attack data packets have a time to live value (TTL) of 126 or 253, whilst the packets of the network traffic mostly have a TTL of 127 or 254. However, TTL values of 126 and 253 do not happen in the training vectors of the attack types (McHugh, 2000). Second, the probability distribution of the testing set is different from the probability distribution of the training set, because of inserting new attack vectors in the testing set (Mahoney & Chan, 2003; Vasudevan, Harshini, & Selvakumar, 2011). This leads to skew or bias classification methods towards some records rather than balance between the attack and normal vectors. Third, the data set is out-dated; as a result, it does not a comprehensive representation of contemporary normal and attack vectors (Tavallaee, Bagheri, Lu, & Ghorbani, 2009).

## DESCRIPTION OF THE UNSW-NB15 DATA SET

The UNSW-NB 15 ("UNSW-NB15 data set," 2015) data set was developed by using an IXIA tool to extract a hybrid of modern normal and modern attack behaviors. This data set involves nine attack categories and 49 features (Moustafa & Slay, 2015b). This data set contains 2, 540,044 observations. A part of this data set was divided into training and testing sets, reflected in Table 2.

*Table 2: A Part of the UNSW-NB15 Data Set Distribution*

| Category | Training set | Testing set |
|---|---|---|
| Normal | 56,000 | 37,000 |
| Analysis | 2,000 | 677 |
| Backdoor | 1,746 | 583 |
| DoS | 12,264 | 4089 |
| **Category** | **Training set** | **Testing set** |

| Exploits | 33,393 | 11,132 |
| Fuzzers | 18,184 | 6,062 |
| Generic | 40,000 | 18,871 |
| Reconnaissance | 10,491 | 3,496 |
| Shellcode | 1,133 | 378 |
| Worms | 130 | 44 |
| **Total Records** | **175,341** | **82,332** |

The UNSW-NB15 data set has several advantages when compared to the NSLKDD data set. First, it contains real modern normal behaviors and contemporary synthesised attack activities. Second, the probability distribution of the training and testing sets are similar. Third, it involves a set of features from the payload and header of packets to reflect the network packets efficiently. Finally, the complexity of evaluating the UNSW-NB15 on existing classification systems showed that this data set has complex patterns. This means that the data set can be used to evaluate the existing and novel classification methods in an effective and reliable manner.

## PROPOSED ARCHITECTURE FOR ADAPTIVE NIDS

In this section, we describe the architecture of adopting the relevant features for each class, whether normal or abnormal, choose the training and testing sets, execute some classification methods as a decision engine and evaluate the outcome of the decision engine. Figure 1 represents the procedures of applying the architecture to execute an adaptive NIDS in a very short time processing as follows:
- Choose an input data set, for example UNSW-NB15 or NSLKDD data set.
- Execute Algorithm 1 to compute the Central Points (CP) of attribute values.
- The CP output is the input of Algorithm 2 to calculate the high ranked attributes.
- Divide the data set into two parts: training set and testing set to learn classifiers.
- Apply EM clustering, Naïve Bayes and Logistic regression techniques as decision engines.
- Evaluate the performance in terms of accuracy and FAR, with respect to the processing time.

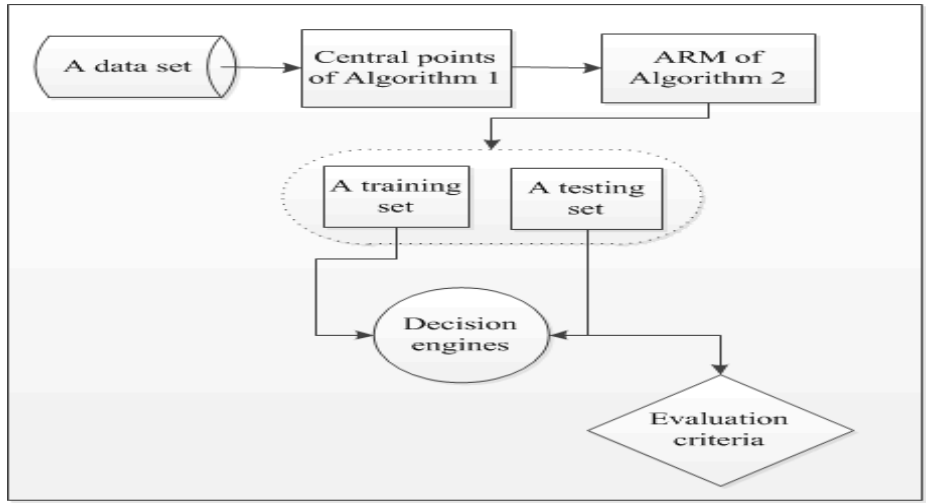

*Figure 1: The Proposed Architecture for an Adaptive NIDS*

**Central Points of Attribute Values**

To reduce the processing time, the data set records are divided into equal segmentations using Equation 1. The purpose of the data set partitioning is to be easier during the processing and identify statistical characteristics, for example mean or mode, from different parts of records. This leads to accomplish the reliability of results by adopting the relevant attributes.

$$p = \# \ of \ partions = \frac{\# \ of \ records}{\# \ of \ attributes} \quad (1)$$

In each part of the data set, we compute the mode which is the most frequent value of a feature (Runnenburg, 1978). The attribute values of a network data set could be numeric or categorical, as in the below example.

*An example to compute the mode of attribute values*

1. **Numeric values**
   X={1, 2, 1, 1, 3.2, 1}          > **mode** = {1}
   ----------------------------------------------------
2. **Categorical values**
   Y={'tcp', 'udp', 'tcp', 'udp'}   > **mode** = {'udp'}

In Algorithm 1, the central points of attribute values (mode) are described. In line 1 and 2, the for loops assign all data values. From line 3 to 12, check attribute values either categorical or numerical, and then compute the mode for each data part (p). Lines 13 to 17 repeat the steps until finishing all parts. Line 18 retrieves the mode of all data parts to be input for computing the ARM.

```
Algorithm 1: Central Points Of Attribute Values
Input: d data set, p
1.  for (r=1 to length(row)) do
2.      for (c=1 to length(col)) do
3.          if (d[r][c] != categorical) then
4.              pre[r][c] = mode(d_{1:p})
5.              if (pre[r][c] != 0 ) then
6.                  centres [r][c]= + pre[r][c]
7.              end if
8.          else
9.              pre[r][c] = count (d_{1:p})
10.             If ( pre[r][c] > pre[r][c+1]) then
11.                 centres [r][c]= + pre[r][c]
12.             end if
13.             p = p - 1
14.             row=row - (row/p)
15.         end if
16.     end for
17. end for
18. return centres
```

**Feature Selection ARM**

An ARM (Agrawal et al., 1993; Ma, 1998) is a data mining method to compute the correlation of two or more than two attributes in a data set, because it can find the strongest itemsets between observations. To explain the ARM, let $r = \{f_1, f_2, f_3, \ldots, f_N\}$ be a set of features and D be a data set consisting of T transactions $t_1, t_2, t_3, \ldots, t_N$. Each transaction $t_j$, $\forall\ 1 \leq j \leq N$ is a set of features such that $t_j \subseteq r$. The association rule ($f_1 (i.e., antecdent) \Rightarrow f_2 (i.e., precedent)$) subjects to the constraints of (1)$\exists\ t_j, f_1, f_2 \in t_j$, (2)$f_1 \subseteq r, f_2 \subseteq r$, and (3)$f_1 \cap f_2 \in \emptyset$.

The ARM subjects to two methods: *support* and *confidence* to create rules. *Support* determines the frequency of row values that denotes the association percentage, as reflected in Equation (2). In Equation (3), *confidence* is the frequency of a precedent if the antecedent has already occurred.

$$sup\ (f_1 \Rightarrow f_2) = \frac{|\#t_j | f_1, f_2 \in t_j|}{N} \quad (2)$$

$$conf\ (f_1 \Rightarrow f_2) = \frac{|\#t_j | f_1, f_2 \in t_j|}{|\# t_j | f_1 \in t_j|} \quad (3)$$

The ARM finds out all repeated itemsets and identifies the strongest rules in the frequent itemsets. The strongest ARM in D is realised, if the *support* of a rule is greater than a user-specified minimum support ($sup \geq minsup$), and confidence of a rule is greater than minimum confidence thresholds ($conf \geq minconf$).

It is clear that the CP of attribute values of Algorithm 1 is considered as an input of Algorithm 2 to reduce the processing time. Algorithm 2 generates the highest ranked attributes based on the ARM. Line 1 is a loop to all CP. From line 2 to 14, check if the rules do not accomplish the ARM constraints, remove it. Otherwise, compute support and confidence. In Line 15, all rules order descending based on the values of support and confidence. From Line 17 to 21, the strongest features are selected based on the number of required features.

```
Algorithm 2: Feature Selection Based On The ARM
Input:   centres (C), minimum support (minsup), label (L)
         minimum confidence (L), number of required features (X)
Output:  F (feature subset)
1.  for (i=1 to length(C)) do
2.      if (C[i]==C[i+1])  then
3.          count[i] = count[i] + 1
4.      else
5.          count[i]=1
6.      end if
7.      filter_C[i]= C- C[i]
8.  end for
9.  for (j=1 to length(filter_C)) do
10.     if (count[j]<=1) then
11.         sup[j] = count[j] / length(filter_C)
12.         conf[j] = count[j] / length (D[j])
13.     end if
14. end for
15. Sort (filter_C, sup, conf)
16. for (m=1 to X) do
17.     If (sup >= minsup && conf>= minconf)  then
18.         F + =extracted_features (r, L)
19.     end If
20. end For
21. return F
```

**Decision Engine Techniques**

In the Decision engine, we used Expectation-Maximisation (EM) clustering (Bradley, Fayyad, & Reina, 1998), Logistic Regression (LR) (Kleinbaum & Klein, 2010) and Naïve Bayes (NB) (Panda & Patra, 2007) techniques. First, The EM technique maximises the probability density function of a Gaussian distribution to compute the mean and the covariance of each attribute in a data set. The EM clustering includes two steps: Expectation (E-step) and Maximization (M). In the E-step, the likelihood of observation is calculated, while the M-step re-estimates the parameter values from the E-step to accomplish the highest expected outcome. Second, The NB is a conditional probability model which creates the classification of the two classes: normal (0) or attack (1). It is computed using the maximum a posterior, as denoted as:

$$P(L|I) = \underset{w \in \{1,2,...,N\}}{\operatorname{argmax}} P(L_w) \prod_{j=1}^{N} P(I_j|L_w) \quad (4)$$

such that $L$ denotes the label, $I$ is the observation of each class, $w$ is the class number, $P(L|I)$ refers to the probability of the class given a specified observation and $\prod_{j=1}^{N} P(I_j|L_w)$ is multiplication of all the probabilities of the instances conditionally to their classes to achieve the maximum outcome. Third, the LR algorithm constructs the correlation between a dependent variable (*L*) and independent variables (*F*). It utilises the maximum likelihood function to estimate the regression parameters, as in Equation (4).

**Evaluation Criteria**

The classification measures are four elements: $TP, TN, FP$ and $FN$. First, $TP$ (true positive) is the number of correctly classified attacks. Second, $TN$ (true negative) is the number of correctly classified normal records. Third, $FP$ (false positive) is the number of misclassified attacks. Finally, $FN$ (false negative) denotes the number of misclassified normal records. The accuracy is the percentage of the correctly classified records over all the rows of data set, whether correctly or incorrectly classified(Moustafa & Slay, 2015a), as reflected in the following Equation:

$$acc = \frac{TP + TN}{TP + TN + FP + FN} \quad (5)$$

The False Alarm Rate (FAR) reflects the rate of the misclassified to classified records, as in Equation (8). Equations (6) and (7) calculate False Positive Rate (FPR) and the False Negative Rate (FNR), respectively.

$$FPR = \frac{FP}{FP + TN} \quad (6)$$

$$FNR = \frac{FN}{FN + TP} \quad (7)$$

$$FAR = \frac{FPR + FNR}{2} \quad (8)$$

Precision and recall are computed, as in Equations (9) (10). The precision is the fraction of correctly classified attacks to all attack records. On the other hand, the recall is the fraction of correctly classified attacks to the number of correctly classified attacks and misclassified attacks.

$$precision = \frac{tp}{tp + fp} \quad (9)$$

$$recall = \frac{tp}{tp + fn} \quad (10)$$

The Equations (5) to (8) evaluate the efficiency and reliability of the Decision engine. It is acknowledged that the highest trusted detection is accomplished, when the accuracy value closes to 100% and FAR closes to 0%.

## EXPERIMENTAL RESULTS AND DISCUSSION

The hybrid feature selection is developed using Visual studio C# 2008. Each data set divided into equal parts using Equation 1. The UNSW-NB15 data set involves 5601 partitions. In contrast, the NSLKDD involves 3072 partitions. Figure 2 represents the construction of the central points of each attribute values, with respect to the processing time. The parts of the UNSW-NB15 data set consumed about 2 minutes to calculate the most frequent values for each feature. Conversely, the parts of the NSLKDD data set consumed about 1.4 minutes. This CP algorithm iteratively generates the highest repeated values to reduce the cost of generating support and confidence of the ARM.

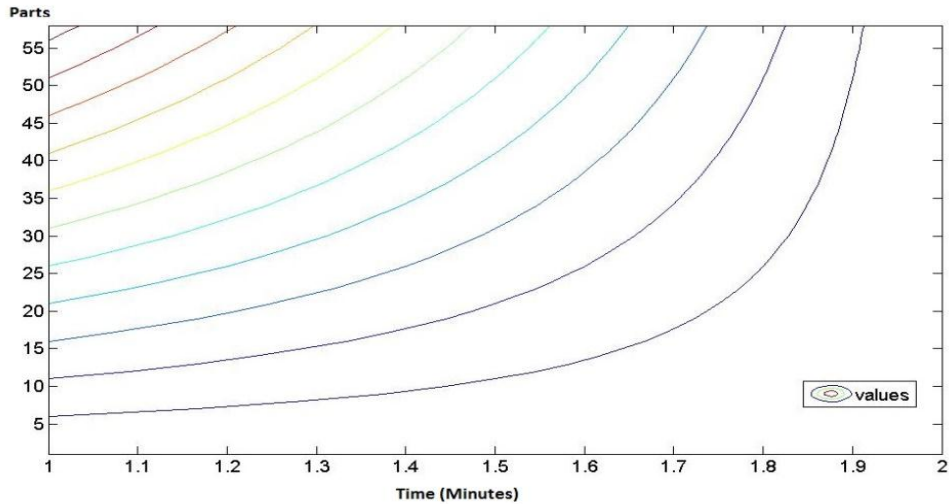

*Figure 2: Data Set Segmentations vs. Processing Time*

Algorithm 2 is executed with three different values of *minsup* and *minconf*: 0.4, 0.6 and 0.8 to estimate the reliability of outcomes. The goal of selecting these values is that the results of probability could be low (0 - 0.4), median (0.4 - 0.6) or high (0.61- 1). Therefore, we consider that the low, median and high probabilities may be up to 0.4, 0.6 and 0.8 in average, respectively. We select some rules based on these probabilities to generate the most important rules. In Figure 3, some rules and their importance are elicited to choose a set of features for each class either normal (0) and attack (1). The importance of rules equals the average of support and confidence of a rule.

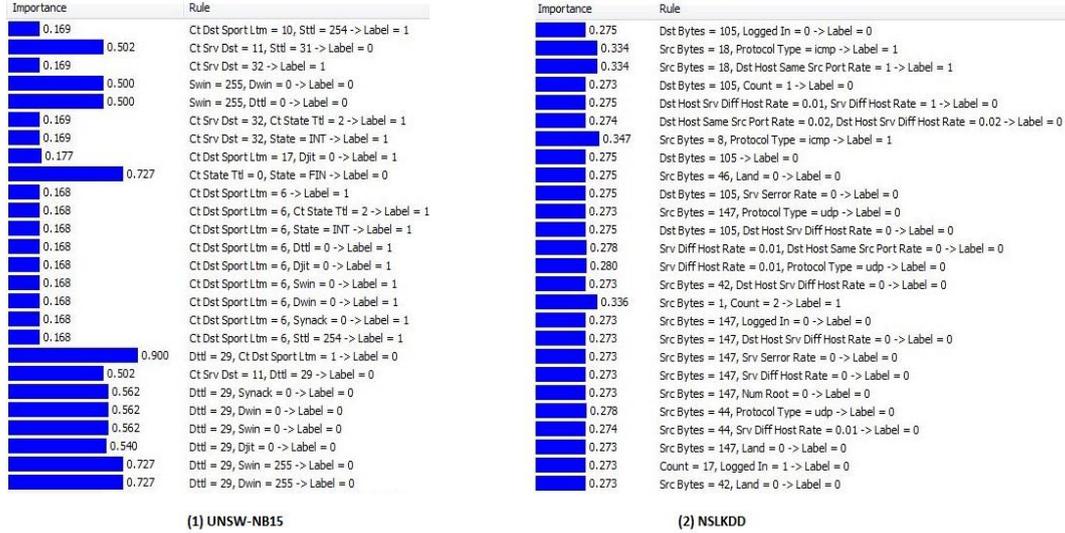

*Figure 3: A part Of The ARM On The Two Data Sets*

Finally, the highest generated attributes from the association rule are ranked. We adopted the highest 11 attributes to ensure at least 25% of all features will be used in the decision engine, as reflected in Table 3. We attempted to select less than 11 attributes, the evaluation of decision engine techniques were extremely unsatisfactory. Therefore, it can be observed that 25% of features consider a reliable outcome because all features are independent and identically distributed (i.i.d.). Table 3 includes the selected 11 features for the UNSW-NB15 and the NSLKDD data sets.

*Table 3: The Adopted Features of the Two Data Sets*

| UNSW-NB15 Features | NSLKDD Features |
|---|---|
| state | dst_bytes |
| Dttl | dst_host_srv_diff_host_rate |
| synack | srv_diff_host_rate |
| swin | land |
| dwin | dst_host_same_src_port_rate |
| ct_state_ttl | count |
| ct_src_ltm | src_bytes |
| ct_srv_dst | logged_in |
| Sttl | protocol_type |
| ct_dst_sport_ltm | num_root |
| Djit | srv_rerror_rate |

The evaluation criteria of executing the Expectation-Maximisation clustering (EM), the Logistic Regression (LR) and the Naïve Bayes (NB) are calculated in terms of accuracy and False Alarm Rate (FAR) to evaluate the complexity of these data sets, as referred in Table 4.

*Table 4: The Evaluation of the Two Data Sets*

|  | UNSW-NB15 | | NSLKDD | |
|---|---|---|---|---|
|  | Accuracy | FAR | Accuracy | FAR |
| EM | 77.2 | 13.1 | 74.4 | 14.2 |
| LR | 83.0 | 14.2 | 82.1 | 17.5 |
| NB | 79.5 | 23.5 | 28.9 | 61.4 |

The LR produces the best results on the two data sets. On the contrary, the NB reflects the worst outcome on the two data sets. In a general overview, the accuracy and FAR of these techniques on the UNSW-NB15 attributes are better than the NSLKDD attributes. Further, in the techniques, the precision and recall on the UNSW-NB15 data set are higher than on the NSLKDD data set, as shown in Figure 4.

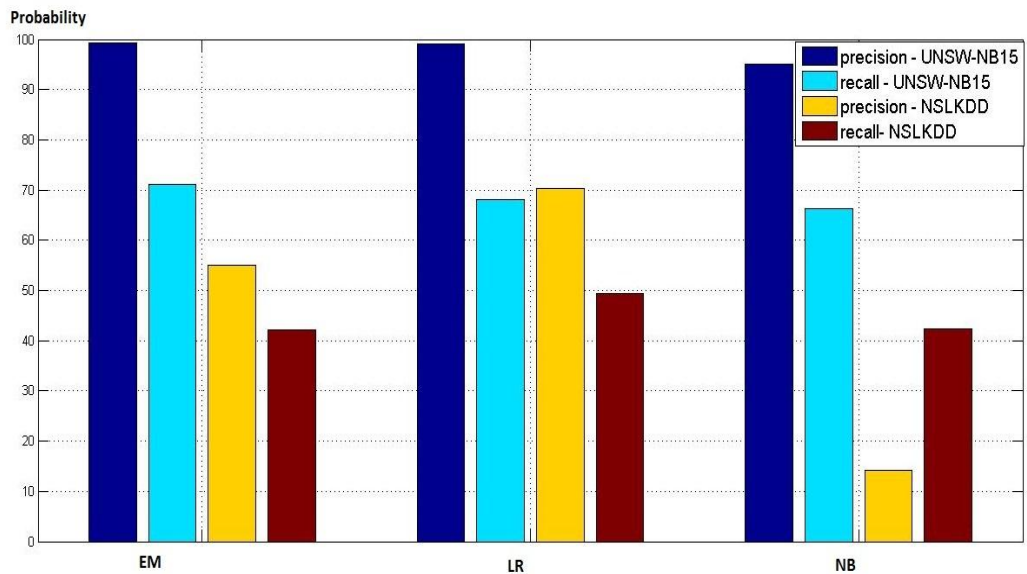

*Figure 4: Precision and Recall of the Two Data Sets*

To clarify why UNSW-NB15 data set has a better assessment than the NSLKDD, there are two main reasons. First, the NSLKDD data set has many features have the majority of value '0' whether in the observations of normal or abnormal, so the decision engine techniques are not able to distinguish between normal and attack observations. However, the UNSW-NB15 has a wide variety of values that represent the nature of modern real network in which these values are similar for attack and normal records. Second, the data distribution of NSLKDD data set in the training and testing set is different, due to the addition of new attacks to the testing set. Conversely, the data distribution of UNSW-NB15 data set in the training and testing set is similar, because of all observations were generated from a same test-bed.

## CONCLUSION AND FUTURE WORK

In this paper, we propose a hybrid feature selection technique based on the central points (CP) of attribute values and Association Rule Mining (ARM). The CP helps to reduce the processing time overall by selecting the most frequent values. The ARM is customised to choose the highest ranked features by removing irrelevant or noisy features. This algorithm is executed on the UNSW-NB15 and the NSLKDD data set. To discriminate between attack and normal records, the EM clustering, Logistic Regression and Naïve Bayes are used. The results show that, the evaluation of the UNSW-NB15 data set is better than the NSLKDD. Ultimately, the proposed feature selection technique has two advantages: reduce processing time and improve the evaluation of decision engines.
In the future, we plan to enhance the algorithm by adding steps that help in reducing the processing time of similar values.